\documentclass[a4paper,11pt]{article}
\pdfoutput=1 
\usepackage{jheppub} 
\usepackage[T1]{fontenc} 
\usepackage{amsmath,amssymb,epsfig,txfonts}
\usepackage{xspace}

\def\etmiss{\ensuremath{E_{T}^{\mathrm{miss}}}\xspace}

\def\ptmiss{\ensuremath{\vec p^{\mathrm{\ miss}}_T\xspace}}

\newcommand{\mttwo}{\ensuremath{m_{T2}}\xspace}

\newcommand{\ttbarZ}{\ensuremath{t\bar{t}Z}\xspace}
\newcommand{\ttbarDM}{\ensuremath{t\bar{t}+E_{T}^{\mathrm{miss}}}\xspace}
\newcommand{\tWDM}{\ensuremath{tW+E_{T}^{\mathrm{miss}}}\xspace}

\newcommand{\ttbar}{\ensuremath{t\bar{t}}\xspace}

\newcommand{\bjet}{\ensuremath{b}-jet\xspace}
\newcommand{\bjets}{\ensuremath{b}-jets\xspace}
\newcommand{\btagged}{\ensuremath{b}-tagged\xspace}

\newcommand{\mblvar}{\ensuremath{m_{b l}^{t}}\xspace}
\def\TeV{\ifmmode {\mathrm{\ Te\kern -0.1em V}}\else
                   \textrm{Te\kern -0.1em V}\fi}
\def\GeV{\ifmmode {\mathrm{\ Ge\kern -0.1em V}}\else
                   \textrm{Ge\kern -0.1em V}\fi}

\def\bm#1{\mbox{\boldmath$#1$\unboldmath}}
\def \beq{\begin{equation}}
\def \eeq{\end{equation}}
\def \bea{\begin{eqnarray}}
\def \eea{\end{eqnarray}}

\begin{document}

\title{Searching for production of dark matter \\ in association with  top quarks at the LHC}

\author[1]{Ulrich Haisch}

\author[2]{and Giacomo Polesello}

\affiliation[1]{Max Planck Institute for Physics, F{\"o}hringer Ring 6,  80805 M{\"u}nchen, Germany}

\affiliation[2]{INFN, Sezione di Pavia, Via Bassi 6, 27100 Pavia, Italy}

\emailAdd{haisch@mpp.mpg.de}
\emailAdd{giacomo.polesello@cern.ch}

\abstract{In the framework of spin-0 $s$-channel dark matter (DM)  simplified models, we reassess the sensitivity of future LHC runs to the production of DM in association with top quarks. We consider two different missing  transverse energy ($E_T^{\mathrm{miss}}$) signatures, namely production of DM in association  with either a $t \bar t$ pair or  a top quark and a $W$ boson, where the latter channel has not been the focus of a dedicated analysis prior to this work. Final states with two leptons are studied and a realistic  analysis strategy is developed that simultaneously takes into account both channels. Compared to other existing search strategies the proposed  combination of $t \bar t + E_T^{\mathrm{miss}}$ and $t W + E_T^{\mathrm{miss}}$ production  provides a significantly  improved coverage of the parameter space of spin-0 $s$-channel DM simplified models. }

\maketitle
\flushbottom

\section{Introduction}

In models where dark matter~(DM) is pair-produced via an $s$-channel process~(see~e.g.~\cite{Abdallah:2015ter,Abercrombie:2015wmb,Boveia:2016mrp}) the corresponding mediators have to be  singlets under both the QCD and QED part of the standard model~(SM) gauge group. Gauge invariance then implies that pure vector or axialvector mediators have to couple  flavour universally to the SM quarks, an option that is severely constrained by the existing LHC searches for dijet and/or dilepton resonances. The couplings of scalar and pseudoscalars mediators to quarks are instead restricted by  the stringent bounds on flavour changing neutral current processes. The simplest way to avoid the latter limits consists in invoking the  minimal flavour violation (MFV) hypothesis~\cite{D'Ambrosio:2002ex}. In the spin-0 case, this hypothesis implies that the couplings of the mediators to SM matter are proportional to the fermion masses. One is thus naturally led to consider scalar and pseudoscalar mediators that couple most strongly to the third generation. Similar to the SM Higgs boson, such states can be produced at the LHC through loop-induced gluon fusion or in association with top (or bottom) quarks before they decay to the heaviest kinematically allowed SM final state or DM. Drell-Yan production instead provides generically only very weak  constraints on MFV   spin-0 $s$-channel   DM simplified models. 

In this work, we  reassess the sensitivity of future LHC runs to the production of DM in association with top quarks in the framework of spin-0 $s$-channel  DM simplified models. We focus on missing transverse energy ($\etmiss$) signatures with two leptons in the final state, and thus consider the two channels \ttbarDM and \tWDM. While the former process has received quite some attention (see~\cite{Buckley:2014fba,Haisch:2015ioa,Backovic:2015soa,Buckley:2015ctj,Arina:2016cqj,Haisch:2016gry} for recent theoretical studies), single-top processes involving significant amounts of~$\etmiss$ have been considered only by a few authors. In the publication~\cite{Pinna:2017tay} it has  first been pointed out that production of DM in association with a single top can be used to increase the reach of \ttbarDM analyses (see very recently also~\cite{CMS-PAS-EXO-18-010,ATL-PHYS-PUB-2018-036}). The work~\cite{Plehn:2017bys} has then  shown that a dedicated search for $t$-channel $tj + \etmiss$ production can provide an interesting stand-alone DM signal at the~LHC. Both of these studies have been performed in the context of spin-0 $s$-channel   DM simplified models. The article~\cite{Pani:2017qyd} has finally studied single-top plus $\etmiss$ production in an ultraviolet~(UV) complete model with  two Higgs doublets~(2HDM) and an extra pseudoscalar mediator, known as the 2HDM+a model~\cite{Ipek:2014gua,No:2015xqa,Goncalves:2016iyg,Bauer:2017ota,Tunney:2017yfp,Abe:2018bpo}. In the framework of the 2HDM+a model it has been shown that \tWDM is the most interesting channel because it can be resonantly procedure  via the exchange of a charged Higgs boson. A~dedicated analysis of the \tWDM final state in  spin-0 $s$-channel  DM simplified models has instead not been performed so far. The main goal of this work  is to close this gap by developing an analysis which provides an  improved coverage of the parameter space of spin-0 $s$-channel  DM simplified models via a  combined  two-lepton  search for \ttbarDM and \tWDM production. 

The outline of this article is as follows. In Section~\ref{sec:DMsimp} we describe  the structure of the~DM simplified models that we  consider. Section~\ref{eq:unitarity} contains a discussion of unitarity violation in DM plus top-quark production. From this discussion one can conclude that calculations of the \tWDM signature in the context of spin-0 $s$-channel  DM simplified models lead to meaningful results at LHC energies. The angular correlations of the \tWDM signal are discussed in Section~\ref{sec:anatomy}, while a brief description of our Monte Carlo (MC) and detector simulations is presented in Section~\ref{sec:MC}.  Our actual analysis strategy can be found in Section~\ref{sec:strategy}. It  spells out all selection criteria and  illustrates their impact on the DM  signals and the SM background. In~Section~\ref{sec:results}, we present the numerical results of our analysis, providing a detailed evaluation of the achievable sensitivity  of    \ttbarDM and \tWDM production at future LHC runs.  Constraints on the parameter space of the spin-0 $s$-channel  DM simplified models are also derived in this section. We conclude in Section~\ref{sec:conclusions}. 

\section{Preliminaries}
\label{sec:DMsimp}

The spin-0 $s$-channel  DM simplified models that we are considering in our article can be described by the following interactions (see for instance~\cite{Abdallah:2015ter,Abercrombie:2015wmb,Boveia:2016mrp})
\beq \label{eq:lagrangians}
\begin{split}
{\cal L}_\phi & \supset -g_\chi \hspace{0.25mm} \phi  \hspace{0.25mm}  \bar \chi \chi - \frac{\phi}{\sqrt{2}} \sum_{q=u,d,s,c,b,t}  g_q  \hspace{0.25mm}  y_q  \hspace{0.25mm} \bar q q \,, \\[2mm]
{\cal L}_a & \supset -i g_\chi  \hspace{0.25mm} a  \hspace{0.25mm}  \bar \chi  \hspace{0.25mm} \gamma_5  \hspace{0.25mm} \chi - i  \hspace{0.25mm} \frac{a}{\sqrt{2}} \sum_{q=u,d,s,c,b,t}  g_q  \hspace{0.25mm}  y_q  \hspace{0.25mm} \bar q  \hspace{0.25mm} \gamma_5  \hspace{0.25mm} q \,.
\end{split}
\eeq
Here $\phi$ ($a$) is a scalar (pseudoscalar), $\chi$ represents the DM particle assumed to be a Dirac fermion, $g_\chi$ is a dark-sector Yukawa coupling,  $y_q = \sqrt{2} m_q/v$ are the SM quark Yukawa couplings with~$m_q$ the mass of the relevant quark $q$ and $v \simeq 246 \, {\rm GeV}$ the Higgs vacuum expectation value, and $\gamma_5$ finally denotes the fifth Dirac matrix. 

\begin{figure}[t!]
\begin{center}
\includegraphics[width=0.95\textwidth]{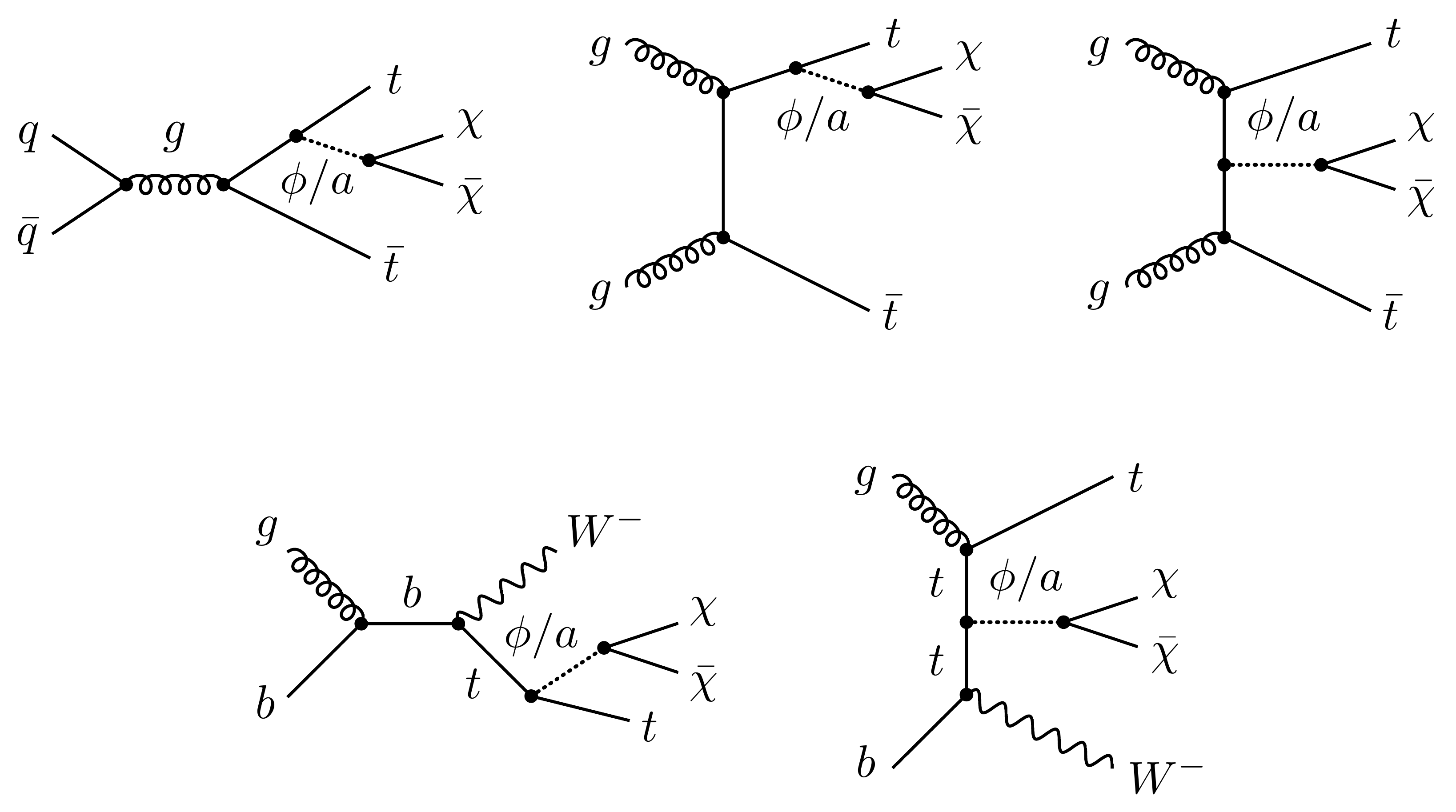}
\vspace{2mm}
\caption{Representative Feynman diagrams for \ttbarDM~(upper~row) and \tWDM~(lower~row) production that involve the exchange of a scalar ($\phi$) or pseudoscalar ($a$) mediator with the interactions specified in~(\ref{eq:lagrangians}). }
\label{fig:diagrams}
\end{center}
\end{figure}

The Lagrangians ${\cal L}_{\phi/a}$  are not invariant under the SM gauge group, meaning that additional particles and/or  interactions need to be invoked to embed~(\ref{eq:lagrangians})  into a gauge-invariant structure at the electroweak scale. Scalar and pseudoscalar couplings of the form~(\ref{eq:lagrangians}) do for example arise in 2HDM extensions with an extra spin-0 gauge singlet~\cite{Ipek:2014gua,No:2015xqa,Goncalves:2016iyg,Bell:2016ekl,Bauer:2017ota,Tunney:2017yfp,Abe:2018bpo}. In fact, in models of this type there is a well-defined limit, known as alignment with decoupling, in which the Lagrangians~${\cal L}_{\phi/a}$  contain all the interactions necessary to predict the \ttbarDM and \tWDM signatures up to terms that formally vanish  for infinitely heavy 2HDM  spin-0 states~$H, A, H^\pm$.  There are thus well-motivated~UV complete scenarios for which the above interactions  provide the dominant contributions  to \ttbarDM and \tWDM production (see~Figure~\ref{fig:diagrams} for example Feynman diagrams). As a result~(\ref{eq:lagrangians}) can also serve as a useful starting point for further phenomenological investigations of the interplay and complementarity of the  \ttbarDM and \tWDM channels in certain limits of next-generation spin-0 DM models. 

\section{Unitarity considerations}
\label{eq:unitarity}

An issue related to the discussion presented at the end of the last section is whether  the production of DM in association with top quarks can be reliable calculated in the context of~(\ref{eq:lagrangians}). To address this question, we will study the UV~behaviour of  scattering amplitudes (cf.~\cite{Bell:2015sza,Kahlhoefer:2015bea,Bell:2015rdw,Haisch:2016usn,Englert:2016joy} for related discussions). Since the \ttbarDM amplitudes arising  from~${\cal L}_{\phi/a}$ turn out to be well-behaved in the high-energy limit, the ensuing discussion will entirely focus on the   \tWDM signature. 

The Feynman graphs that lead to \tWDM production in the spin-0 $s$-channel DM simplified models~(\ref{eq:lagrangians}) are shown in the lower row of Figure~\ref{fig:diagrams}. Diagrams where the mediator~$\phi/a$ is radiated off a bottom quark are suppressed by the bottom Yukawa coupling and will be neglected in the further discussion. In the high-energy limit, the emission of the $W$ boson can be treated in the effective $W$-boson approximation~\cite{Dawson:1984gx,Kane:1984bb}. In this approximation, the process $g b \to t W^- \chi \bar \chi$ factorises into the interactions $g b \to b$ (or $g t \to t$), the hard scattering $b \to t W^- \phi/a$ and the decay $\phi/a \to \chi \bar \chi$.  This factorisation allows one to study the high-energy behaviour of the process $b \to t W^- \phi/a$ to gain an understanding of the energy growth of the full $g b \to t W^- \chi \bar \chi$ amplitude. Our discussion will follow closely the arguments given in~\cite{Maltoni:2001hu,Farina:2012xp}. 

In the high-energy limit, where the Mandelstam variables fulfil $\hat s, - \hat t, -\hat u \gg m_t^2, m_W^2, m_{\phi}^2$ and for a longitudinal-polarised $W$ boson, the $b \to t W^- \phi$  amplitude takes the following form in the scalar $s$-channel DM simplified model:
\beq \label{eq:Aphi}
{\cal A}_\phi= V_{tb}^\ast \, \frac{y_t}{v} \, g_t \, \bar b \hspace{0.25mm} P_L \hspace{0.25mm} t \,.
\eeq
Here $V_{tb}$ is the relevant Cabibbo-Kobayashi-Maskawa matrix element and $P_{L} = (1-\gamma_5)/2$ projects onto left-handed  fields. In the case of the pseudoscalar mediator $a$, one instead finds~${\cal A}_a = - i {\cal A}_\phi$. 

\begin{figure}[t!]
\begin{center}
\includegraphics[width=0.975\textwidth]{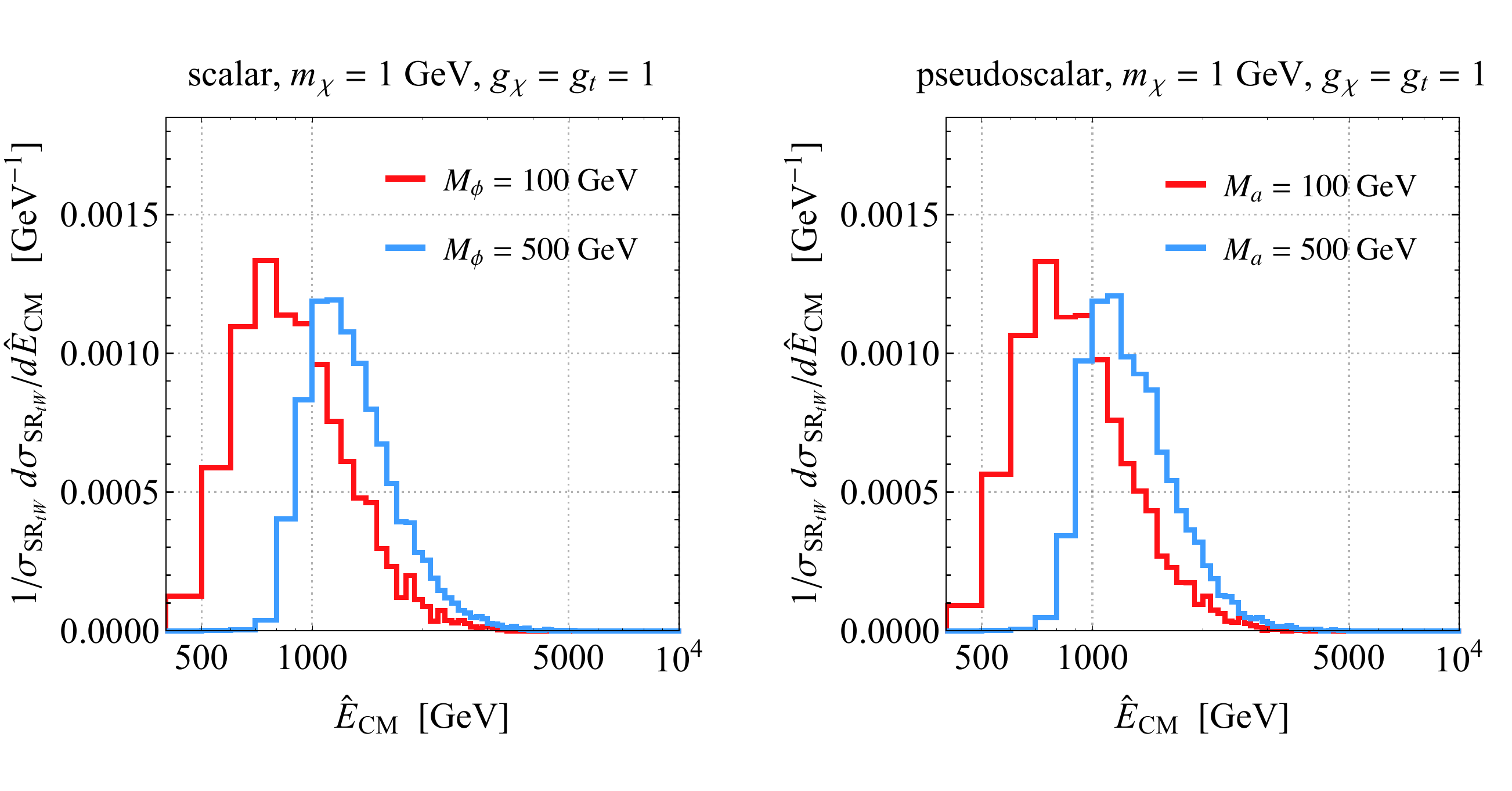}
\vspace{-5mm}
\caption{Normalised differential fiducial cross sections for $\tWDM$ production as a function of the partonic CM energy in the scalar~(left) and pseudoscalar~(right) $s$-channel  DM simplified model. The shown predictions correspond to $14 \, {\rm TeV}$ $pp$ collisions, the selections ${\rm SR}_{tW}$ defined in Table~\ref{tab:srdef} and the  parameter choices  indicated in the legends and the headlines of the two panels.}
\label{fig:unitarity}
\end{center}
\end{figure}

The above results imply that the $s$-wave amplitude $a_0$ of the $b \to t W^- \phi/a$ process grows linearly with the partonic centre-of-mass (CM) energy $\hat E_{\rm CM} = \sqrt{\hat s}$. Explicitly, we find that 
\begin{equation} \label{eq:a0}
|a_0| = \frac{|V_{tb}^\ast|}{24 \hspace{0.25mm} \pi}  \hspace{0.25mm} \frac{y_t}{v} \; g_t  \hspace{0.5mm} \hat E_{\rm CM} \,,
\end{equation}
which allows to estimate the cut-off scale $\Lambda$ where perturbative unitarity is lost.  Imposing the condition $|a_0|  < 1$ and identifying $\hat E_{\rm CM} \simeq \Lambda$, the relation~(\ref{eq:a0}) leads to  
\begin{equation} \label{eq:Lambda}
\Lambda \simeq \frac{24 \hspace{0.25mm} \pi}{ |V_{tb}^\ast |}  \hspace{0.25mm}  \frac{v}{ y_t} \, \frac{1}{g_t}  \simeq \frac{18.6 \, {\rm TeV}}{g_t} \,,
\end{equation}
where to obtain the final result we have used that $V_{tb} \simeq 1$ and $y_t \simeq 1$. In order to make the $g b \to t W^- \chi \bar \chi$ amplitudes well-behaved additional particles and/or couplings thus have to appear at a scale around~$\Lambda$, allowing to embed~(\ref{eq:lagrangians})  into a gauge-invariant UV complete theory. 

 While the question to which extent  these new states and/or interactions will modify the DM phenomenology is a model-dependent issue, the relation~(\ref{eq:Lambda}) can be used to determine whether the \tWDM signature can be reliably computed in the framework of spin-0  $s$-channel  DM simplified models~(\ref{eq:lagrangians}). To do this, we  study the  dependence of the fiducial  \tWDM  cross sections on the partonic CM energy, employing the experimental selection that  is used in Section~\ref{sec:strategy} to disentangle the   \tWDM  signal from the SM background.  The corresponding signal region is called ${\rm SR}_{tW}$ and the relevant cuts can be found in Table~\ref{tab:srdef}. 

In  Figure~\ref{fig:unitarity} we present normalised distributions for scalar~(left) and pseudoscalar~(right) mediators of type~(\ref{eq:lagrangians}) obtained with the MC setup described in Section~\ref{sec:MC}.   The shown predictions correspond to  $pp$ collisions at $14 \, {\rm TeV}$ and the   ATLAS/CMS DM Forum~(DMF)~\cite{Abercrombie:2015wmb,Boveia:2016mrp} benchmark parameters $m_\chi = 1 \, {\rm GeV}$ and $g_\chi = g_t = 1$. One observes that for a mediator mass of $100 \, {\rm GeV}$~($500 \, {\rm GeV}$) the spectra of the normalised fiducial cross sections have  peaks at around  $700 \, {\rm GeV}$~($1.1  \, {\rm TeV}$) and tails that rapidly fall off for increasing~$\hat E_{\rm CM}$. Numerically, we find for example that the phase-space region with $\hat E_{\rm CM} > 3 \, {\rm TeV}$ contributes only a fraction of $0.2\%$ to $1.2\%$ to the  fiducial \tWDM cross sections for the model realisations considered in the figure. The fraction of  \tWDM events with  partonic CM energies at or above the cut-off~(\ref{eq:Lambda})  is hence negligible at 14 TeV. In~consequence, the perturbative computation of the \tWDM signal can be fully trusted,  and unitarity violation in the production of DM in association with top quarks  is spurious at LHC  energies in the spin-0  $s$-channel  DM simplified models~(\ref{eq:lagrangians}). 

\section{Angular correlations of the \bm{\tWDM} signal}
\label{sec:anatomy}

In the case of the \ttbarDM  signal, it has been shown in~\cite{Buckley:2015ctj,Haisch:2016gry} that information on the spin-0 mediator type  is encoded in the correlations between the final-state top quarks as well as their decay products. As explained in the publication~\cite{Haisch:2016gry}, the observed angular correlations in \ttbarDM  production can be understood from the  energy behaviour of the fragmentation function $f_{t \to \phi/a}$ that describes the radiation of $\phi/a$ fields from top quarks. In the case of the interactions~(\ref{eq:lagrangians}),  the leading (universal) fragmentation functions read~\cite{Dawson:1997im, Dittmaier:2000tc}
\beq \label{eq:ffs}
\begin{split}
f_{t \to \phi}  = \frac{g_t^2}{(4 \pi)^2} \left [ \frac{4 \left (1 -x \right )}{x} + x \ln \left ( \frac{\hat s}{m_t^2} \right ) \right ] \,, \qquad f_{t \to a}   = \frac{g_t^2}{(4 \pi)^2} \left [  x \ln \left ( \frac{\hat s}{m_t^2} \right ) \right ] \,, 
\end{split}
\eeq
where it is assumed  that $\hat s \gg 4 m_t^2 \gg M_{\phi/a}^2$  and $\sqrt{\hat s} = 2 \hat E/x$ with $\hat E$ denoting the energy of the emitted mediator. The first expression  in~(\ref{eq:ffs}) implies  that  a light scalar is radiated off top quarks preferentially with small energy (or equivalent small momentum fraction~$x$) due to the soft singularity proportional to~$1/x$ in $f_{t \to \phi}$. This soft-enhanced term is instead absent in the pseudoscalar fragmentation function $f_{t \to a}$. 

\begin{figure}[t!]
\begin{center}
\includegraphics[width=0.975\textwidth]{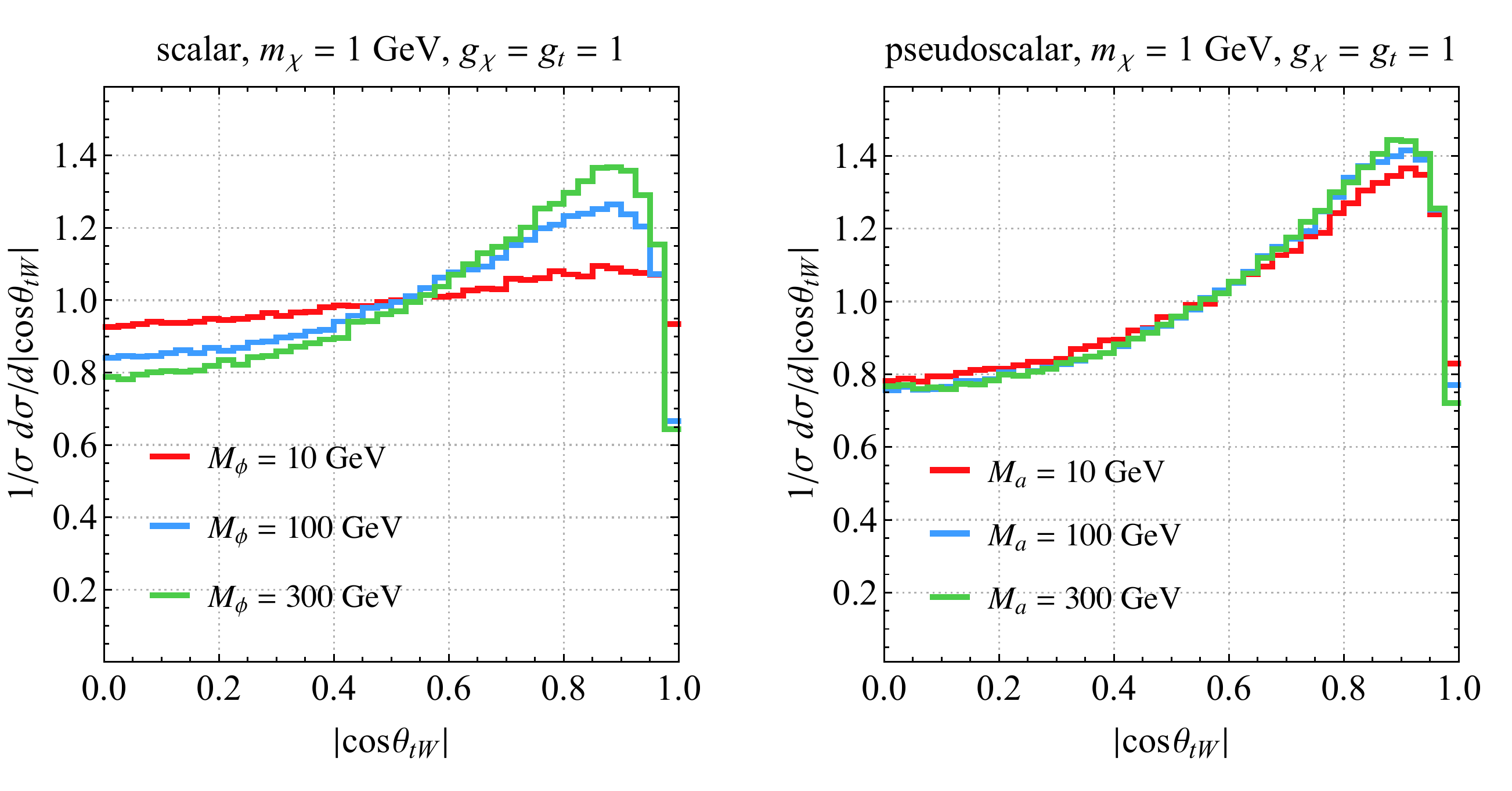}
\vspace{-2mm}
\caption{Normalised $| \hspace{-0.4mm}\cos \theta_{tW}|$  distributions for the $\tWDM$ signature in the scalar~(left) and pseudoscalar~(right) $s$-channel  DM simplified model. The shown results correspond to the 14 TeV LHC and the parameters given in the legends and headlines of the figure. }
\label{fig:angular1}
\end{center}
\end{figure}

The arguments presented in~\cite{Haisch:2016gry} can also be used to gain a qualitative understanding of  the angular correlations of the \tWDM signal (see  also \cite{181111048} for a related discussion). Below we will focus on $\cos \theta_{tW} = \tanh \left ( \Delta \eta_{tW}/2 \right )$, because in the work~\cite{Haisch:2016gry} the related observable $\cos \theta_{t \bar t}$  has been shown to be a good CP analyser  in the case of the $\ttbarDM$ signature. Here $\Delta \eta_{tW}$ denotes the   difference in pseudorapidity of the top quark and the $W$ boson. All~spectra shown in this subsection have been obtained with the MC setup described in Section~\ref{sec:MC}, correspond to $pp$ collisions at $14 \, {\rm TeV}$ and employ the DMF benchmark parameter choices $m_\chi = 1 \, {\rm GeV}$ and $g_\chi = g_t = 1$. We emphasise that the shapes of the angular correlations discussed below as well as in Section~\ref{sec:strategy} are independent of the specific choices made for $m_\chi$, $g_\chi$ and $g_t$. This feature can be understood by noting that the two  considered DM signatures  factorise into  $pp \to t \bar t + \phi/a$ or $pp \to t W + \phi/a$ production followed by $\phi/a \to \chi \bar \chi$. As a result of this factorisation the normalised kinematic distributions of the $t \bar t$ or $t W$ systems as well as their decay products  do only depend on the mass $M_{\phi/a}$ relevant for on-shell production and decay but not on the other spin-0 $s$-channel DM simplified model parameters. We finally mention that we have  explicitly verified the parameter independence of the normalised angular correlations of the $\ttbarDM$ and \tWDM signal by generating MC events using $M_{\phi/a} = 300 \, {\rm GeV}$,  $g_\chi = g_t = 1$ and $m_\chi = 100 \, {\rm GeV}$ instead of  $m_\chi = 1 \, {\rm GeV}$.

In Figure~\ref{fig:angular1} we  present results for the $| \hspace{-0.4mm} \cos \theta_{tW}|$  distributions in the \tWDM channel for three different realisations of the scalar and pseudoscalar  $s$-channel  DM simplified models~(\ref{eq:lagrangians}). The~displayed spectra  are all normalised to unity and no selection cuts have been imposed. One sees that for a very light mediator of $10 \, {\rm GeV}$~(red curves)  the  normalised~$| \hspace{-0.4mm} \cos \theta_{tW}|$ spectrum is almost flat in the scalar case, while it is enhanced toward larger values of $| \hspace{-0.4mm} \cos \theta_{tW}|$ for a pseudoscalar. The flatness of the scalar spectrum is a result of the $1/x$ singularity of the fragmentation function~$f_{t \to \phi} $~$\big ($cf.~(\ref{eq:ffs})$\big)$, which favours emissions of soft~$\phi$ fields. Compared to a pseudoscalar, a scalar  of the same mass hence tends to be produced more forward, and in consequence the accompanying top quark and $W$ boson are produced more central. On average the observable $| \hspace{-0.4mm} \cos \theta_{tW}|$ is thus smaller  for a very light scalar than a pseudoscalar. For heavier mediators of  $100 \, {\rm GeV}$~(blue curves) the soft enhancement of  $f_{t \to \phi}$ is less important, and the $| \hspace{-0.4mm} \cos \theta_{tW}|$ spectra all develop a peak close to~1 irrespectively of the mediator type. The peaks become more pronounced and the shapes more similar when the mediator mass is increased further to $300 \, {\rm GeV}$~(green curves). \\

\begin{figure}[t!]
\begin{center}
\includegraphics[width=0.975\textwidth]{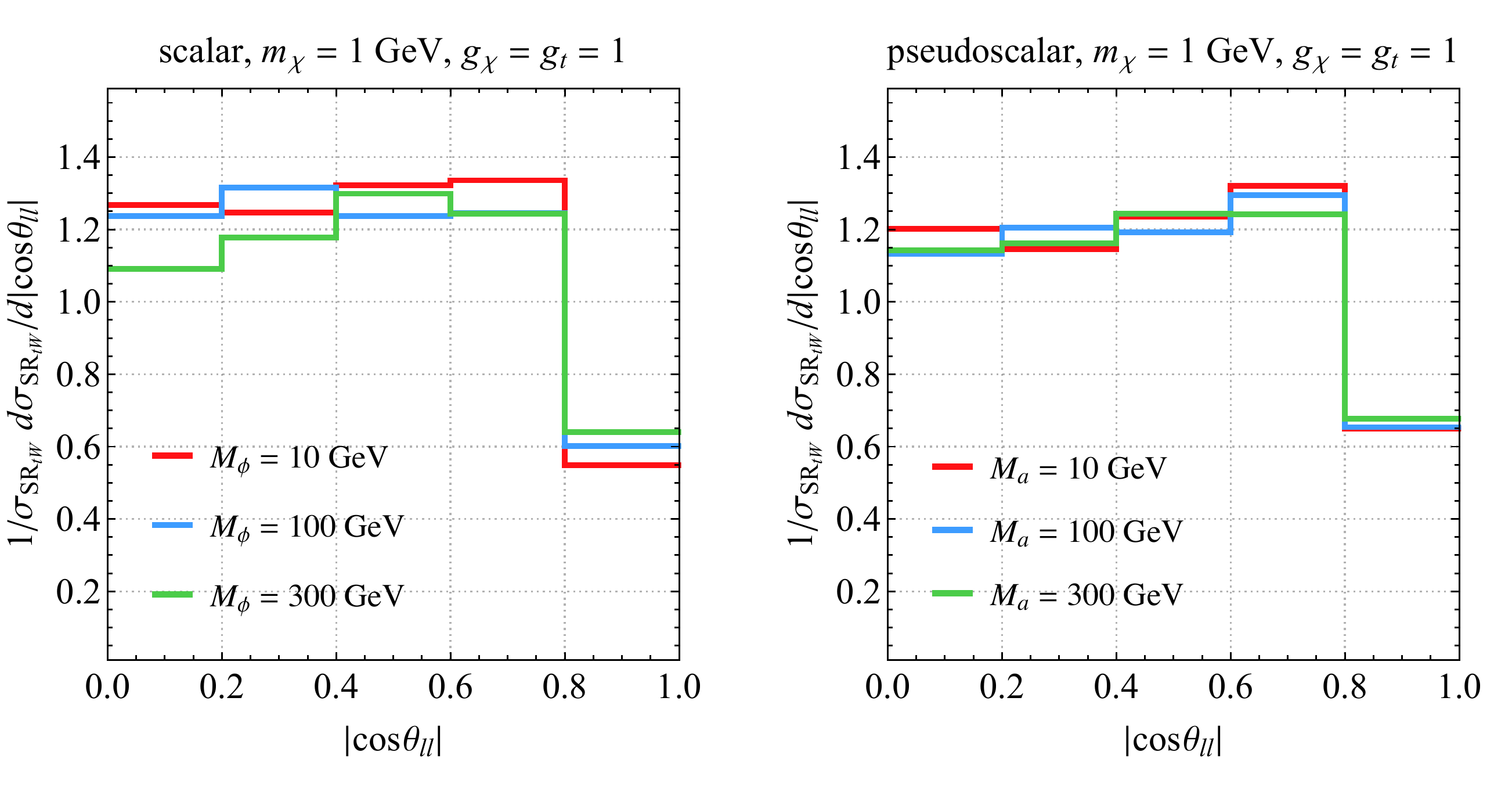}
\vspace{-2mm}
\caption{Normalised $|\hspace{-0.4mm} \cos\theta_{ll}|$ distributions corresponding to $\tWDM$ production with subsequent decays $t \to bW \to bl\nu$ and $W \to l \nu$. The shown results correspond to $pp$ collisions at 14 TeV and adopt apart from $C_{\rm em} > 150 \, {\rm GeV}$ the experimental selections ${\rm SR}_{tW}$ as given in~Table~\ref{tab:srdef}. The parameter choices of the scalar~(left) and pseudoscalar~(right) $s$-channel  DM simplified models are indicated in the panels.}
\label{fig:cthrec}
\end{center}
\end{figure}

The results presented above suggest  that for light spin-0 mediators angular correlations in the~$tW$ system can be used to study the CP properties of the mediation mechanism in \tWDM production.  The directions of the  top quark and the $W$ boson are, however, experimentally not directly accessible, since the dilepton final state contains  four invisible particles (i.e.~two neutrinos and two DM particles). The relative orientation of the top quark and the $W$ boson has hence to be obtained indirectly from measurements of the angular distributions of the charged-lepton pairs resulting from  $t \to bW \to b l \nu$ and $W \to l \nu$. The crucial question is whether the discriminating power of angular variables such as $\cos\theta_{ll}$   survives the experimental cuts necessary to extract the DM signal from the SM backgrounds. To address this issue we show in Figure~\ref{fig:cthrec} normalised $|\hspace{-0.4mm} \cos\theta_{ll}|$ distributions of the \tWDM signature, imposing apart from $C_{\rm em} > 150 \, {\rm GeV}$ the selection requirements~SR$_{tW}$ as specified in  Table~\ref{tab:srdef}. We again display results for the three mediator masses $10 \, {\rm GeV}$, $100 \, {\rm GeV}$ and $300 \, {\rm GeV}$. One observes that the signal selections tend to wash out  shape differences in the distributions of the $|\hspace{-0.4mm} \cos\theta_{ll}|$ variable, thereby reducing the sensitivity of this observable in realistic experimental analyses of the \tWDM signal. Compared to the  \ttbarDM  case analysed in detail in the work~\cite{Haisch:2016gry}, angular correlations of the \tWDM signal alone thus turn out to be  less useful as a model discriminant.  However, we will see later  that the observable~$|\hspace{-0.4mm} \cos\theta_{ll}|$ still provides  sensitivity to the mass and the CP nature of the mediating particle if the dilepton events resulting from the \ttbarDM and \tWDM channels are combined into a single signal sample as done in~Section~\ref{sec:strategy}.

\section{MC generation and detector simulation}
\label{sec:MC}

All the  signal samples shown in this article are generated at next-to-leading order (NLO) using  a slightly modified version of the {\tt DMsimp} implementation~\cite{Backovic:2015soa} of the Lagrangians~(\ref{eq:lagrangians}) together with {\tt  MadGraph5\_aMC@NLO}~\cite{Alwall:2014hca} and {\tt NNPDF3.0} parton distribution functions (PDFs)~\cite{Ball:2014uwa}. The final-state top quarks and $W$ bosons are decayed  with {\tt MadSpin}~\cite{Artoisenet:2012st} and the events are showered with {\tt PYTHIA~8.2} \cite{Sjostrand:2014zea}. We produce both the \ttbarDM and the \tWDM final states assuming $pp$ collisions at 14 TeV. Following~\cite{Demartin:2016axk,Haisch:2018djm},  we avoid double counting in the \tWDM channel by removing all doubly resonant diagrams involving top quarks, employing the so-called diagram removal procedure~\cite{Frixione:2008yi}. We consider nine different values of the mediator masses $M_{\phi/a}$, varying from $10 \, {\rm GeV}$ to $1\, {\rm TeV}$.  The mass of the DM particles is set to $m_\chi = 1 \, {\rm GeV}$ and we employ $g_\chi = g_t =1$ for the mediator couplings  to DM and to top quarks. Given the strong Yukawa suppression the mediator couplings to the bottom quark and all other light fermions are set to zero. This simplification has no impact on the results obtained in our work. The widths of the mediators are assumed to be minimal and calculated at tree level using {\tt  MadGraph5\_aMC@NLO}. Since in the narrow width approximation the signal predictions factorise into the cross sections for $pp \to \ttbar/tW + \phi/a$ production times the $\phi/a \to \chi \bar \chi$ branching ratio, changing the mediator width leads only  to a rescaling of the signal strength. The experimental acceptance is instead insensitive to the mediator width, and therefore it is sufficient to generate samples for a single  coupling choice.  The predictions for other coupling values can then be simply obtained by scaling with the corresponding $\phi/a \to \chi \bar \chi$ branching ratio that dictates the fraction of invisible decays of the mediators.  

To model the $\ttbarDM$ and $\tWDM$ backgrounds accurately,  SM processes involving at least two leptons coming from the decay of  vector bosons are generated. Backgrounds either with fake electrons ($e$) from jet misidentification or with real non-isolated leptons  from the decay of heavy-flavoured hadrons are not considered in our analysis. A reliable estimate of these backgrounds depends on a detailed simulation of the detector effect beyond the scope of this paper. This simplification is motivated by the results of the ATLAS  analysis \cite{Aaboud:2017rzf}, which addresses kinematic configurations similar to the ones targeted in the present study, and finds that  the background from non-prompt leptons does not exceed the level of $15\%$ of the total background. The backgrounds from~$\ttbar$~\cite{Campbell:2014kua}, $tW$~\cite{Re:2010bp}, $WW$, $WZ$ and $ZZ$ production~\cite{Melia:2011tj,Nason:2013ydw} were all generate at NLO with {\tt POWHEG~BOX} \cite{Alioli:2010xd}. The $Z+{\rm jets}$ samples are generated at leading order~(LO) with {\tt  MadGraph5\_aMC@NLO} and contain up to four jets. {\tt  MadGraph5\_aMC@NLO} is also used to simulate the $\ttbar V$ backgrounds with $V = W,Z$ at LO with a multiplicity of up to two jets. All partonic events are showered with {\tt PYTHIA~8.2}. The samples produced with {\tt POWHEG~BOX} are normalised to the NLO cross section given by the generator, except $t\bar{t}$ which is normalised to the  cross section obtained at next-to-next-to-leading order (NNLO) plus next-to-next-to-leading logarithmic accuracy~\cite{Czakon:2011xx,Czakon:2013goa}. The $V+{\rm jets}$ samples are normalised to the known NNLO cross sections~\cite{Anastasiou:2003ds,Gavin:2012sy}, while in the case of the~$\ttbar V$~samples  the NLO cross sections calculated with {\tt  MadGraph5\_aMC@NLO} are used as normalisations.  

The actual analysis uses experimentally identified electrons, muons ($\mu$),  photons, jets and $E_{T, \rm miss}$. These objects are constructed from the stable particles in the generator output.  Jets are obtained by clustering the true momenta of all particles but muons that interact in the calorimeters. {\tt FastJet}~\cite{Cacciari:2011ma} is used to construct anti-$k_t$ jets~\cite{Cacciari:2008gp} of radius $R=0.4$. Jets originating from the hadronisation of bottom quarks (\bjets) are experimentally tagged in the detector (\btagged). The variable $\vec p_{T,\rm miss}$ with magnitude $E_{T, \rm miss}$ is defined at truth level,~i.e.~before applying detector effects, as the vector sum of the transverse momenta of all the invisible particles (neutrinos and DM particles in our case). The effect of the detector on the kinematic variables used in the analysis is simulated by smearing with Gaussian functions the momenta of  the  reconstructed objects, and by applying reconstruction and tagging efficiency factors tuned to mimic the performance of the  ATLAS detector~\cite{Aad:2008zzm,Aad:2009wy}. More details on our detector simulation can be found in \cite{Haisch:2016gry,Haisch:2018djm}. 

\section{Analysis strategy}
\label{sec:strategy}

In order to search for the \ttbarDM and \tWDM signatures, we consider  two-lepton final states.  This final state can arise from either  two top-quark decays~($t \to b W \to b l \nu$) or a top-quark decay and a $W$-boson decay~($W \to l \nu$). In both cases, the dominant  backgrounds turn out to be \ttbar and \ttbarZ. Since the amount of \etmiss is typically larger for the signal due to the undetected DM and   the~\mttwo observable~\cite{Lester:1999tx,Barr:2003rg} has a sharp edge at $M_W$ in the case of \ttbar production, these two variables provide the main handles to suppress the leading SM backgrounds. An analysis based on  \etmiss and~\mttwo  has been shown in \cite{Haisch:2016gry,Aaboud:2017rzf} to achieve a good separation of the \ttbarDM signal from the \ttbar~background, whereas a part of  \ttbarZ is irreducible. Applying the same strategy to  \tWDM production would also allow to separate the signal from the \ttbar background, but, due to the lower cross section, it would be difficult to extract the \tWDM signal from the irreducible  \ttbarZ background. Moreover, it would  be complicated to separate  \tWDM  production from the much larger \ttbarDM signal. 

An important ingredient of a realistic analysis strategy thus consists in separating events that are associated to    two semileptonic top-quark decays  from those with one semileptonic top-quark decay and one leptonic $W$-boson decay.  In the former case, the resulting events have two \bjets, while in the latter case only a single \bjet is produced. Vetoing the second \bjet, however, turns out to be an ineffective strategy, as the typical \bjet tagging efficiency of  70\% to 80\% would result in a large surviving \ttbar background. Furthermore, $tW$ production where a bottom quark is extracted from the proton is often accompanied by an additional \bjet, and as a result  this type of SM background cannot be tamed by vetoing the presence of a second \btagged jet in the event. 

Instead of using a $b$-jet veto, our analysis strategy is therefore based on the observation that if a \bjet is produced in the top-quark decay, its invariant mass is bounded from above by $\sqrt{m_t^2 - M_W^2 } \simeq 153 \, {\rm GeV}$. Events  compatible with two semileptonic top-quark decays can thus be selected or rejected by  introducing the observable 
\begin{equation} \label{eq:mblt}
\mblvar =\mathrm{min} \hspace{0.5mm} \Big  (\mathrm{max} \hspace{0.5mm}  \big ( m_{l_1 j_a}, m_{l_2 j_b} \big  ) \Big ) \,,
\end{equation}
where $l_1$ and $l_2$ denote the two final-state leptons and the minimisation runs over all  pairs $\{j_a, j_b\}$ of jets inside a predefined set of test jets. Based on the number of \bjets in the event, the set of test jets is defined as follows. If the event includes one or two \bjets, an additional test jet is considered, chosen as the non-$b$-tagged jet with the highest $b$-tagging weight. If three \bjets are found, they are all taken as test jets. 

\begin{table}
\begin{center}
\begin{tabular}{|l|c|c|}
\hline
 & ${\rm SR}_{t \bar t}$ & ${\rm SR}_{tW}$ \\
\hline
$N_{l}$ & \multicolumn{2}{|c|}{$= 2$, \; \; $p_{T, l_{1}}>25 \, {\rm GeV}$,  \; \; $p_{T, l_{2}}>20 \, {\rm GeV}$, \;\; $|\eta_{l}|<2.5$  }\\
$m_{ll}$  & \multicolumn{2}{|c|}{$>20$~GeV, \; \; $Z$-boson veto for opposite-sign leptons} \\
$N_{b}$ & \multicolumn{2}{|c|}{$>0$, \;\; $p_{T, b}>30 \, {\rm GeV}$, \;\; $|\eta_{b}| < 2.5$}\\
\mttwo  & \multicolumn{2}{|c|}{$>100 \, {\rm GeV}$} \\
\hline
\mblvar & $<160 \, {\rm GeV}$ & $>160$~GeV \; \; || \; \; $N_{j}=1$\\
\hline
$|\Delta\phi_{min}|$ & $>0.8$ & $>0.8$ \\
$|\Delta\phi_{\rm boost}|$ & $<1.2$ & n/a  \\
$M_{\rm scal}$ & n/a & $<500 \, {\rm GeV}$ \\
$C_{\rm em}$ & $> 200 \, {\rm GeV}$ & $> 200 \, {\rm GeV}$ \\
$|\hspace{-0.4mm}\cos\theta_{ll}|$ & shape fit & shape fit \\
\hline
\end{tabular}
\vspace{4mm}
\caption{Definition of the two signal regions ${\rm SR}_{t \bar t}$ and ${\rm SR}_{tW}$. For further details consult the text.}
\label{tab:srdef}
\end{center}
\end{table}

Our search strategy  is based on two distinct signal  regions. A signal region that we call ${\rm SR}_{t \bar t}$, aimed at separating \ttbarDM production from the dominant \ttbar background, and a signal region called ${\rm SR}_{tW}$ that  is designed to disentangling  the \tWDM signal from both \ttbarDM production and the SM background. The definitions of the signal regions are summarised in Table~\ref{tab:srdef}. The first part of the selections is common for the two signal regions. In a first step, events with exactly two isolated oppositely charged  leptons ($N_l = 2$ with  $l = e, \mu$ or one of each flavour) are selected. The individual leptons are required to satisfy $p_{T, l_1} >25 \, {\rm GeV}$, $p_{T, l_2} >20 \, {\rm GeV}$, $|\eta_{l}|<2.5$ and their   invariant mass has to fulfill $m_{ll } > 20 \, {\rm GeV}$.   The kinematic requirements on the charged leptons are such that the events of interest will be selected with full efficiency by the dilepton triggers foreseen for LHC~Run-3 and  the high-luminosity phase of the~LHC~(HL-LHC), as documented in Table~6.4 of~\cite{Collaboration:2285584}. If the charged  leptons are of the same flavour the additional requirement $m_{ll} \notin [71, 111] \, {\rm GeV}$ is imposed to veto events where the charged lepton pair  arises from $Z \to l^+ l^-$.  Furthermore, each event is required to contain at least one \btagged jet ($N_b > 0$) with $p_{T,b}  >30\,{\rm GeV}$ and $|\eta_{b}| < 2.5$. An initial rejection of the SM backgrounds, where~\etmiss is due to the neutrinos from the decay of two $W$ bosons, is finally achieved  by requiring that~$\mttwo>100 \, {\rm GeV}$.

\begin{figure}[t!]
\begin{center}
\includegraphics[width=0.4875\textwidth]{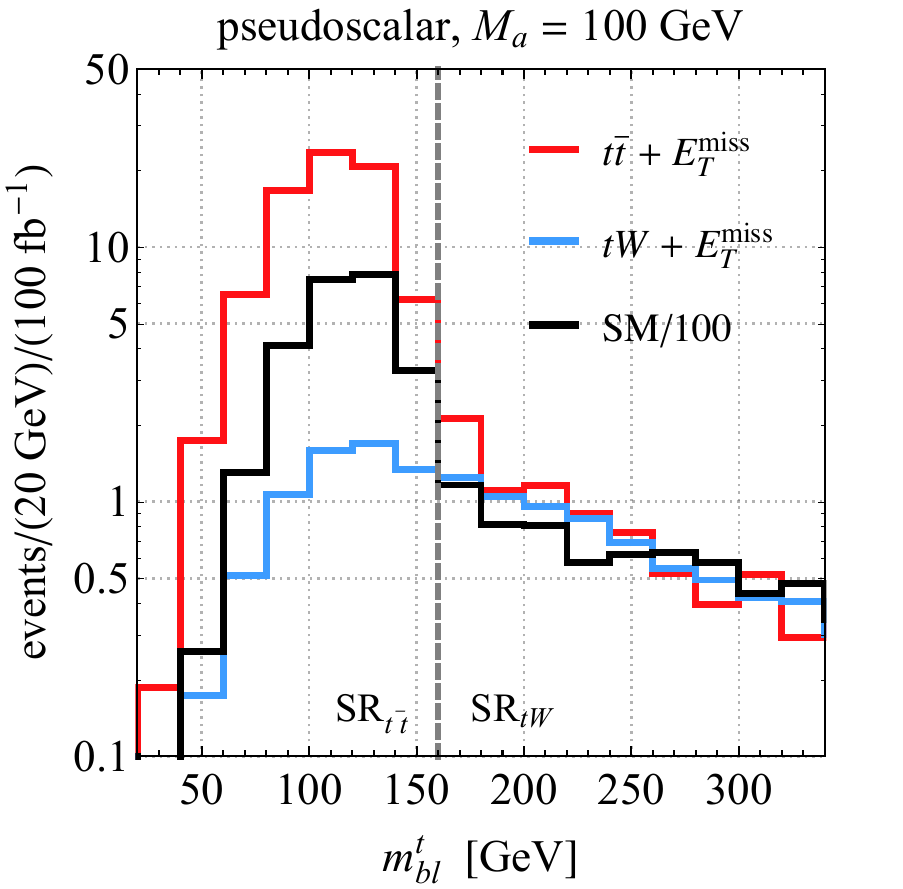}
\vspace{0mm}
\caption{Distributions of the \mblvar variable~(\ref{eq:mblt}) for the \ttbarDM signal,  the \tWDM signal, and the sum of SM backgrounds scaled down by a factor 100. The shown results correspond to $100 \, {\rm fb}^{-1}$ of 14 TeV LHC data and the common lepton, \bjet and \mttwo selections summarised in~Table~\ref{tab:srdef}. The used model parameters are $m_\chi = 1 \, {\rm GeV}$, $g_\chi = g_t =1$ and those appearing   in the headline of the figure. The dashed line corresponds to $m_{bl}^t = 160 \, {\rm GeV}$, which represents the boundary between the two signal regions ${\rm SR}_{\ttbar}$ and ${\rm SR}_{tW}$.}
\label{fig:mblvar}
\end{center}
\end{figure}

The first selection that differs between the two signal regions is a cut on the $\mblvar$ observable introduced in~(\ref{eq:mblt}). In order to illustrate the impact of this selection, we show in Figure~\ref{fig:mblvar} the distributions of $\mblvar$ for the \ttbarDM and the \tWDM signals corresponding to a pseudoscalar of mass $M_a = 100 \, {\rm GeV}$, $m_\chi = 1 \, {\rm GeV}$ and $g_\chi = g_t =1$. The total SM background, which is dominated by \ttbar production, is also displayed. For an easy comparison with the DM signals it has been scaled down by a factor of 100. All event samples correspond to  $100 \, {\rm fb}^{-1}$ of 14 TeV LHC data and the initial selection requirements as described above have been imposed. From the figure it is clear that the region $\mblvar<160 \, {\rm GeV}$ is dominated by the \ttbarDM signal, and is therefore used to define ${\rm SR}_{t \bar t}$.  Numerically, we find for the case at hand that 91\% of the DM signal in ${\rm SR}_{t \bar t}$ arises from  \ttbarDM, while only 9\% are due to \tWDM. In the region $\mblvar>160 \, {\rm GeV}$, one instead obtains similar event fractions of around 50\%  for both \ttbarDM and \tWDM, and hence our signal region ${\rm SR}_{tW}$ adopts this requirement. The variable \mblvar is only defined for events with at least two reconstructed jets with $p_{T,j}  >25\,{\rm GeV}$, and events with only one reconstructed jet ($N_j=1$) are assigned to ${\rm SR}_{tW}$. Relative to the SM background the combined DM signal in  ${\rm SR}_{\ttbar}$~(${\rm SR}_{tW}$) amounts to $3.4\%$  ($2.0\%$). Notice that since the shape of the \mblvar distributions is essentially independent of the type of the mediator, its mass and the other new-physics parameters appearing in~(\ref{eq:lagrangians}), the  $\mblvar$ selection can be applied unchanged for each model realisation. 
 
Further selections are used in the two signal regions to optimise the rejection of the SM backgrounds.  All reconstructed jets with $p_\mathrm{T}^j >25 \, \GeV$ within $|\eta_{\ell}|<2.5$ have to satisfy $|\Delta\phi_{\rm min}|>0.8$ for both signal regions, where~$\Delta\phi_{\rm min}$ corresponds to the angle between $\vec{p}_{\rm T}^{\ j}$ and $\ptmiss$ for the jet closest to $\etmiss$ in the azimuthal plane. The variable $\Delta\phi_{\rm boost}$ defined as the azimuthal angular distance between $\ptmiss$ and the vector sum of $\ptmiss$, $\vec{p}_{T,l_1}$ and $\vec{p}_{T,l_2}$, must satisfy the requirement $|\Delta\phi_{\rm boost}|<1.2$ for ${\rm SR}_{t \bar t}$. In the case of the signal region ${\rm SR}_{tW}$, we in addition require that the scalar sum~$M_{\rm scal}$ of the transverse momenta of all the jets observed in the event satisfies $M_{\rm scal}<500\, {\rm GeV}$. We finally demand that the variable $C_{\rm em} =  \mttwo + 0.2 \, \etmiss$ introduced in~\cite{Haisch:2016gry} exceeds $200 \, {\rm GeV}$ in both signal regions. We will comment further on the impact of this cut  at the end of this section. 

After impose all of these cuts, the surviving SM background amounts to approximately 30 (20) events for ${\rm SR}_{t \bar t}$ (${\rm SR}_{tW}$) per $300 \, {\rm fb}^{-1}$ of integrated luminosity.  The \ttbarDM acceptance in ${\rm SR}_{t \bar t}$ varies between 0.1\% (0.5\%) and 1.9\%~(1.8\%) for scalar (pseudoscalar) mediators with masses between $100 \, {\rm GeV}$ and $1  \, {\rm TeV}$, while the \tWDM acceptance in ${\rm SR}_{tW}$ amounts to between 0.7\%~(1.1\%) and 3.6\%~(3.5\%) for scalar (pseudoscalar) mediators in the same mass range. The fraction of \tWDM events in ${\rm SR}_{t \bar t}$ (${\rm SR}_{tW}$) turns out to be  between 8\%~(50\%) and 20\%~(75\%) for $M_{\phi/a}$ values in the range of $10 \, {\rm GeV}$ to $1 \, {\rm TeV}$, and almost independent of the CP nature of the mediator. These numbers imply that our analysis strategy indeed achieves  the goal of separating the DM signal into two viable signal regions with very different relative contributions from \ttbarDM and \tWDM production. In the range of $10 \, {\rm GeV}$ to $1 \, {\rm TeV}$, the ratio of the number of signal events in ${\rm SR}_{\ttbar}$ and ${\rm SR}_{tW}$ decreases monotonically with mediator mass from around 2.0~(1.5) to 0.8~(1.0) in the scalar (pseudoscalar) case.  After  a LHC discovery, measurements of this ratios together with studies of the~$|\hspace{-0.4mm} \cos\theta_{ll}|$ distribution~\cite{Haisch:2016gry} will therefore provide useful handles to determining the properties of the spin-0 mediators such as mass and CP quantum number. 

\begin{figure}[t!]
\begin{center}
\includegraphics[width=0.975\textwidth]{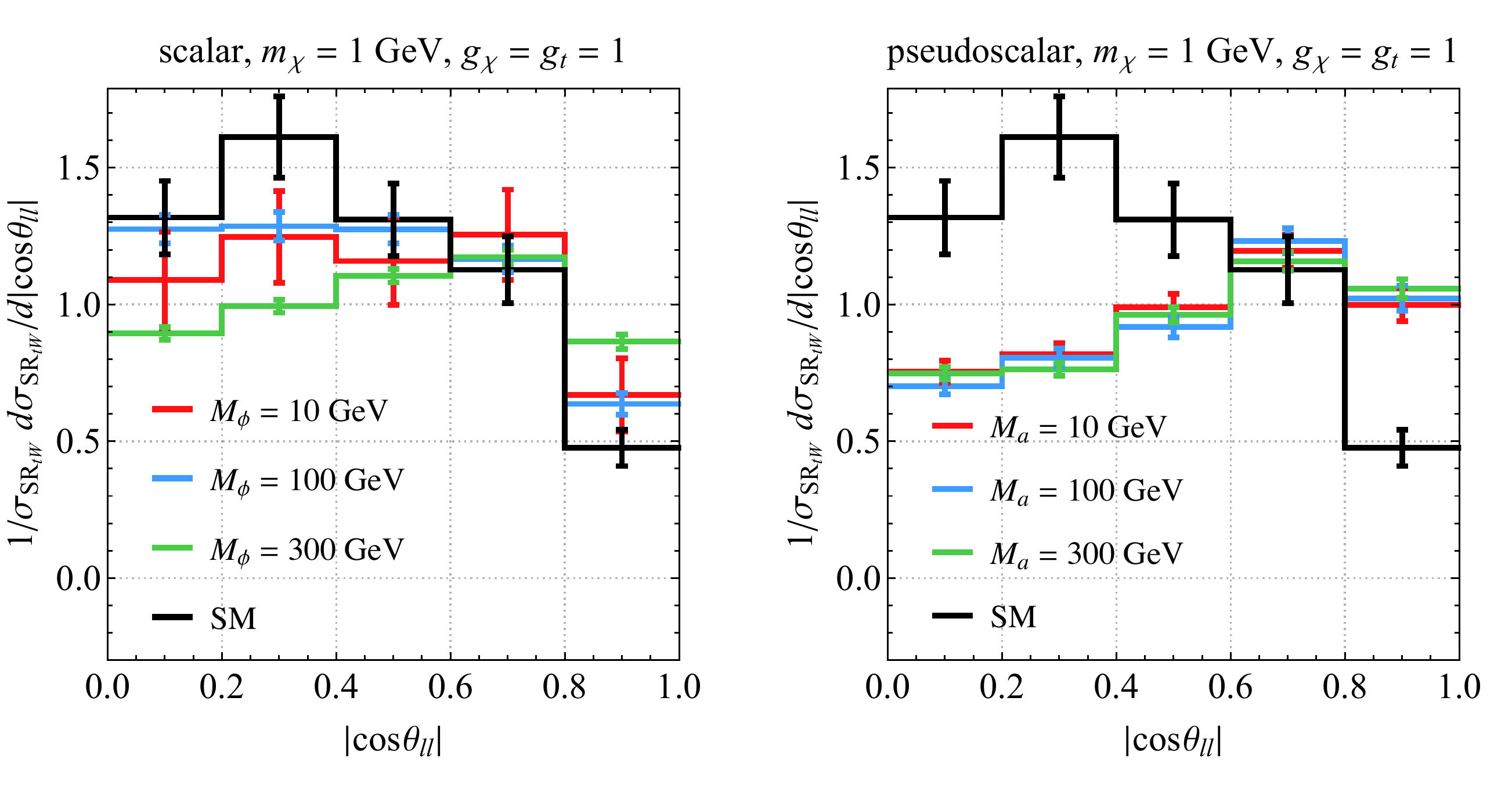}
\vspace{-2mm}
\caption{As Figure~\ref{fig:cthrec} but  combining the dilepton events that arise from both $\ttbarDM$ and  $\tWDM$ production  in spin-0 $s$-channel  DM simplified models into a single sample. The shown error bars represent the statistical errors associated to our MC simulations.  For comparison also the SM distribution is depicted.}
\label{fig:angular3}
\end{center}
\end{figure}

Having developed a realistic strategy for the detection of production of DM plus top quarks,  we can now come back to the issue of angular correlations of the DM signal. Since in the signal region ${\rm SR}_{tW}$ the processes $\ttbarDM$ and $\tWDM$ contribute with similar weight, we show in Figure~\ref{fig:angular3} normalised $|\hspace{-0.4mm} \cos\theta_{ll}|$ distributions of the combined $\ttbarDM$ and  \tWDM signature corresponding, apart from $C_{\rm em} > 150 \, {\rm GeV}$, to the SR$_{tW}$ selection (cf.~Table~\ref{tab:srdef}). Compared to the results presented in Figure~\ref{fig:cthrec}, one observes that the shape differences in the spectra of the $|\hspace{-0.4mm} \cos\theta_{ll}|$ variable are more pronounced once the  dilepton events resulting from $\ttbarDM$ have been added to those coming from $\tWDM$. The observable~$|\hspace{-0.4mm} \cos\theta_{ll}|$ therefore still shows promising discriminating properties in a realistic experimental analysis (such as the one proposed here based on the signal region SR$_{tW}$) where two-lepton events will always arise from the sum  of  $\ttbarDM$  and  $\tWDM$ production. 

\begin{figure}[t!]
\begin{center}
\includegraphics[width=0.4875\textwidth]{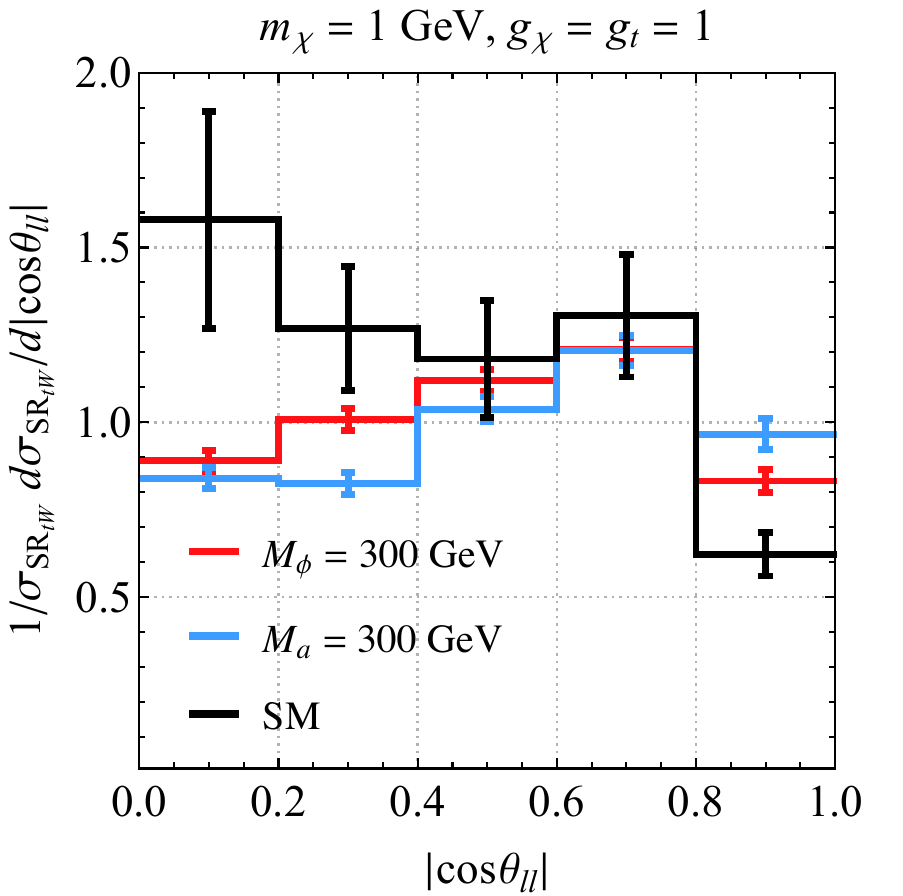}
\vspace{0mm}
\caption{Normalised $|\hspace{-0.4mm} \cos\theta_{ll}|$ distributions of the combined $\ttbarDM$ and $\tWDM$ signal and the SM background  at the 14 TeV LHC. The depicted  results impose the selections ${\rm SR}_{tW}$ summarised in~Table~\ref{tab:srdef}.  The used parameter choices for the $s$-channel  DM simplified models are indicated in the legend and the headline of the panel and the error bars represent the statistical errors associated to our MC simulations.}
\label{fig:angular4}
\end{center}
\end{figure}

To further stress the latter point, we compare in Figure~\ref{fig:angular4} the normalised~$|\hspace{-0.4mm} \cos\theta_{ll}|$ spectra of the SM background  with the combined DM signal for a scalar and pseudoscalar mediator of~$300 \, {\rm GeV}$. The depicted  results correspond to differential  cross sections in the fiducial region~${\rm SR}_{tW}$. The figure demonstrates that  a shape fit to $|\hspace{-0.4mm} \cos\theta_{ll}|$ in the signal region SR$_{tW}$ will enhance the search sensitivity for the combination of $\ttbarDM$  and  $\tWDM$ production. In our actual analysis, we therefore perform  a shape fit to the $|\hspace{-0.4mm} \cos\theta_{ll}|$ observable  in both signal regions. While such a shape~fit has already been employed in~\cite{Haisch:2016gry} for a ${\rm SR}_{\ttbar}\hspace{0.4mm}$-like event selection, in the case of  ${\rm SR}_{tW}$ it is a new ingredient of our search strategy. 

Before presenting the future constraints on the parameter space of the spin-0 $s$-channel DM simplified models, we  add that a shape fit to $C_{\rm em}$ would increase the experimental reach in both signal regions. The obtained improvements depend, however, strongly on the assumed uncertainty on the shape of the sharply falling SM spectrum of $C_{\rm em}$. This uncertainty itself depends on the experimental techniques used to estimate the  backgrounds, and can hence only be determined reliable within the context of an analysis based on real data. We therefore  prefer to use a fixed cut on~$C_{\rm em}$ in both search regions since it is the simplest choice, but still allows to illustrate the interplay of the two signal regions  ${\rm SR}_{\ttbar}$ and ${\rm SR}_{tW}$ in searching for production of DM in association with  top quarks. We finally emphasise that the cut $C_{\rm em} > 200 \, {\rm GeV}$ imposed in both signal regions~(see~Table~\ref{tab:srdef}) is not the optimal choice for all model realisations. For light (heavy) spin-0 $s$-channel mediators a weaker (stronger) $C_{\rm em}$ selection would lead to improved limits. 

\section{Results} 
\label{sec:results}

\begin{figure}[t!]
\begin{center}
\includegraphics[width=0.975\textwidth]{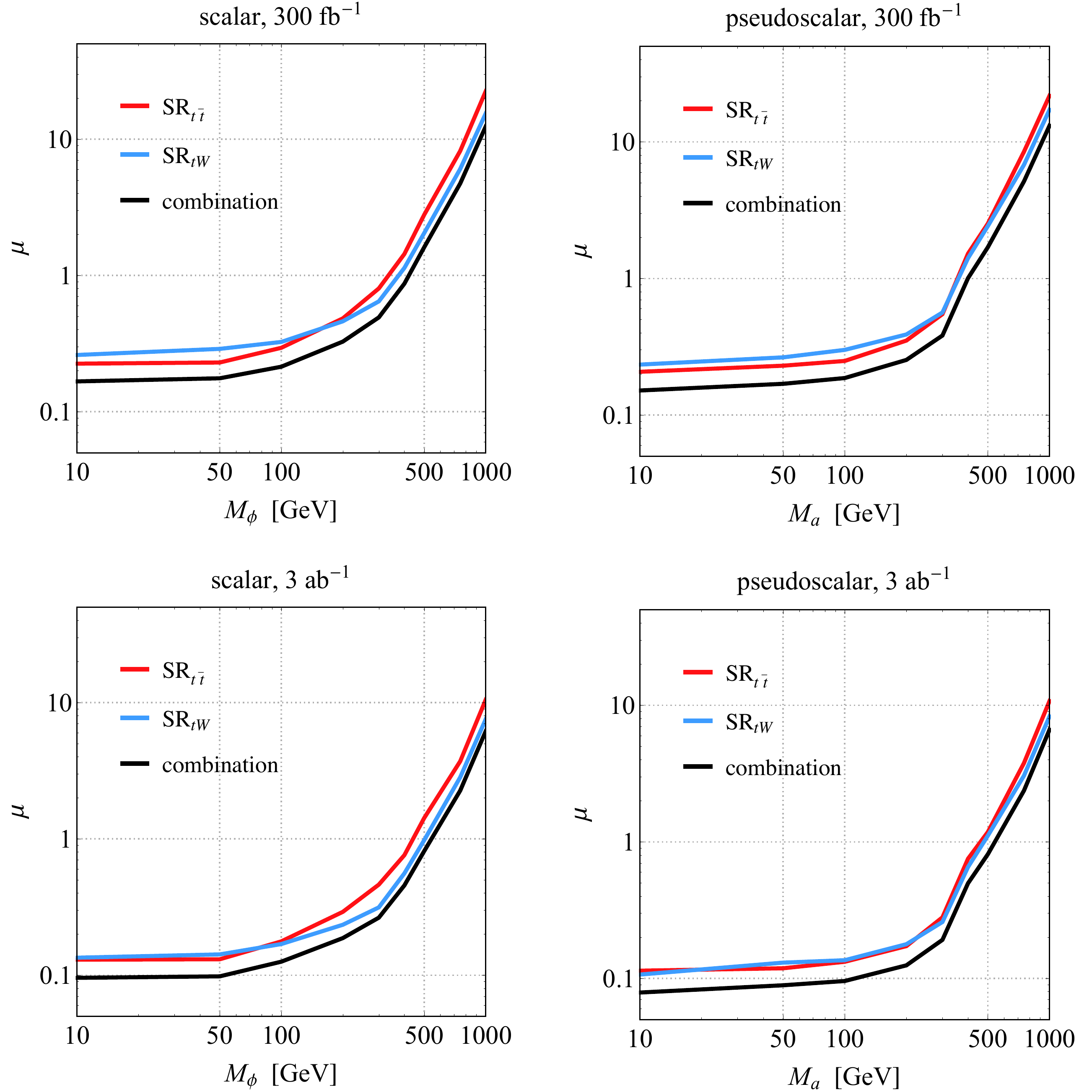}
\vspace{2mm}
\caption{Value of the signal strength $\mu$ that can be excluded at 95\%~CL as a function of the mass for scalar~(left) and pseudoscalar~(right) mediators. The reach for $300 \, {\rm fb}^{-1}$~(upper row) and $3 \, {\rm ab}^{-1}$~(lower row) of 14~TeV~LHC data is shown and the parameter choices $m_\chi = 1 \, {\rm GeV}$ and $g_\chi = g_t = 1$ have been employed. }
\label{fig:reachtw}
\end{center}
\end{figure}

A profiled likelihood ratio test statistic is used to evaluate the upper limit on the ratio of the signal yield to that predicted in the spin-0 $s$-channel  DM simplified models~(\ref{eq:lagrangians}). This ratio will be called signal strength in the following and denoted by~$\mu$. The $C\hspace{-0.2mm}L_s$ method~\cite{Read:2002hq} as implemented in the  {\tt RooStats} toolkit \cite{Moneta:2010pm} is used to derive exclusion limits at 95\%~confidence level~(CL). Figure~\ref{fig:reachtw} displays our limits for integrated luminosities of $300 \, {\rm fb}^{-1}$~(upper row) and~$3 \, {\rm ab}^{-1}$~(lower row) as a function of the mediator mass for scalar~(left) and pseudoscalar~(right) mediators. The standard parameter choices $m_\chi = 1 \, {\rm GeV}$ and $g_\chi = g_t = 1$ are employed. The shown results assume a normalisation uncertainty of 5\% (20\%) for the DM signal~(SM background).  As already  mentioned in Section~\ref{sec:strategy}, in both signal regions a shape fit to the $| \hspace{-0.4mm} \cos\theta_{ll}| $ distribution  is performed using five bins  with a width of 0.2. A bin-by-bin uncertainty of 5\% on the shape of the distribution is assumed. One observes that  the search strategy  ${\rm SR}_{\ttbar}$ tends to perform better than ${\rm SR}_{tW}$ for mediator masses below approximately $150 \, {\rm GeV}$, while for  higher values of $M_{\phi,a}$ the signal region ${\rm SR}_{tW}$  typically provides the more stringent bounds.  Since the two signal regions are designed without overlap, they  can be statistically combined, and the combination yields a notable improvement in reach for both mediator types and over the entire mass range considered.  In fact, our combined analysis leads in the scalar case to the  95\%~CL limit $M_\phi > 417 \, {\rm GeV}$ ($M_\phi > 532 \, {\rm GeV}$)  at $300 \, {\rm fb}^{-1}$~($3 \, {\rm ab}^{-1}$), which  represents a relative improvement of $26\%$ ($22\%$) compared to the  ${\rm SR}_{\ttbar}$ bound.  For pseudoscalar mediators, the corresponding numbers read $M_a > 399 \, {\rm GeV}$ ($M_a > 529 \, {\rm GeV}$)  at $300 \, {\rm fb}^{-1}$ ($3 \, {\rm ab}^{-1}$), implying a relative improvement of $15\%$~($16\%$) with respect to the  ${\rm SR}_{\ttbar}$ analysis alone.   Compared to the earlier studies~\cite{Haisch:2016gry,Pinna:2017tay,Plehn:2017bys,ATL-PHYS-PUB-2018-036} the proposed combination of the \ttbarDM and \tWDM signals hence provides  a significantly improved coverage of the parameter space of  spin-0 $s$-channel  DM simplified models at both LHC~Run-3 and the HL-LHC.

\section{Conclusions}
\label{sec:conclusions}

The goal of this article was  to reassess the future sensitivity of the LHC to the production of DM in association with top quarks  in the framework of  $s$-channel  DM simplified models with scalar and pseudoscalar mediators. We have focused on  final states with large amounts of $\etmiss$ and two charged leptons. Such configurations  can  arise either from  $\ttbarDM$ production followed by two $t \to b W \to b l \nu$ decays or from the $\tWDM$ channel through a $t \to b W \to b l \nu$ decay and a $W \to  l \nu$ decay. Given  the two different production mechanisms leading to the same final state, the two DM signatures generically overlap and should be treated together in an experimental analysis~\cite{Pinna:2017tay,CMS-PAS-EXO-18-010}. 

In order to allow for a theoretically meaningful combination of the $\ttbarDM$ and $\tWDM$ channels in the context of spin-0  $s$-channel  DM simplified models, we have first shown that both signatures can be reliable calculated at LHC energies. To this purpose, we have analysed the  UV~behaviour of the  relevant scattering amplitudes. While the   \ttbarDM amplitude is  well-behaved, the $s$-wave contribution to  \tWDM production is found to diverge linearly with the partonic CM energy. The associated UV cut-off where perturbative unitarity breaks down turns out to be in the multi-TeV range for weakly-coupled models. The fraction of  \tWDM events with such large partonic CM energies is however negligible for $pp$ collisions at 14 TeV (cf.~Figure~\ref{fig:unitarity}).  Unitarity violation in the production of DM in association with top quarks  is therefore spurious at the LHC in  spin-0  $s$-channel  DM simplified models.

We have then studied the angular correlations of the $\tWDM$ signature focusing on the observable $\cos \theta_{tW}$ ($\cos \theta_{ll}$) that measures the difference in pseudorapidity of the top-quark and the $W$ boson (the two charged leptons). In the case of $\cos \theta_{tW}$, we have found that for not too heavy spin-0 particles this variable provides sensitivity to the type of mediation.  The impact of realistic selection cuts on the $\cos \theta_{ll}$ distributions has then been studied, and we observed that  $\cos \theta_{ll}$  looses discriminating power in the signal region ${\rm SR}_{tW}$, specifically designed to search for the  $\tWDM$ signature.   Since in ${\rm SR}_{tW}$ around half of the dilepton events are due to $\tWDM$ and the other half arises from $\ttbarDM$, and given that the $\cos \theta_{ll}$ spectrum of both the  $\ttbarDM$ and the $\tWDM$ 
signals after the ${\rm SR}_{tW}$ selections is significantly different from the SM spectrum (see Figure~\ref{fig:angular4}),  a~shape fit to the $\cos \theta_{ll}$ variable however still turns out to enhance the search sensitivity in practice.

After a brief description of the MC generation and our detector simulation, we have then presented our analysis strategy. Like in the earlier study~\cite{Haisch:2016gry}, the observables  $\etmiss$ and~$\mttwo$  provide the main handles to suppress the SM backgrounds in our work. In order to  separate events that are associated to    two $t \to b W \to b l \nu$ decays  from those with one $t \to b W \to b l \nu$ decay and one $W \to l \nu$ decay, we cut on the $m_{bl}^t$ observable. With this selection we are able to split the search into  two distinct signal regions  (cf.~Figure~\ref{fig:mblvar}). In the first signal region  ${\rm SR}_{\ttbar}$ around 90\% of the dilepton events arise from $\ttbarDM$, while 10\% are due to \tWDM.  In contrast, in the second signal region ${\rm SR}_{tW}$ approximately 50\% of the two-lepton events come from \tWDM and the remaining 50\% originate from \ttbarDM production. 

We have then analysed the coverage of the parameter space of   the spin-0 $s$-channel  DM simplified models expected at LHC~Run-3 and the HL-LHC. For mediator masses below about  $150 \, {\rm GeV}$, we found that a search based on   ${\rm SR}_{\ttbar}$ typically  performs better than one using ${\rm SR}_{tW}$, while for  heavier spin-0 states the opposite tends to be the case. Since the two signal regions ${\rm SR}_{\ttbar}$ and ${\rm SR}_{tW}$ do not overlap, they  can be statistically combined, and it turns out that the combination yields 95\%~CL exclusion bounds on the mediator masses that are improved significantly compared to previous projections~\cite{Haisch:2016gry,Pinna:2017tay,Plehn:2017bys,ATL-PHYS-PUB-2018-036}.  Numerically, we find for the DMF benchmark parameters $m_\chi = 1 \, {\rm GeV}$ and $g_\chi = g_t = 1$, the limits $M_{\phi,a} \gtrsim 410 \, {\rm GeV}$ and $M_{\phi,a} \gtrsim 530 \, {\rm GeV}$ at LHC~Run-3 and the HL-LHC, respectively (see~Figure~\ref{fig:reachtw}).  It should be possible to further strengthen the quoted bounds  by performing a shape fit to $C_{\rm em}$ in the signal regions SR$_{\ttbar}$ and SR$_{tW}$. The obtained improvements depend, however, strongly on the  uncertainty on the shape of the $C_{\rm em}$ distribution within the SM. A~reliable estimate of this uncertainty is only possible in an analysis that uses  real LHC data, and in consequence we have not performed  $C_{\rm em}$ shape fits in this article.  

The analysis strategies proposed by us can also be applied to the  production of DM in association with top quarks  in the framework of next-generation DM simplified models such as for example the 2HDM+a model. A combined study of the \ttbarDM~\cite{Bauer:2017ota} and \tWDM~\cite{Pani:2017qyd} signatures in the context of the 2HDM+a model, while beyond the scope of the work at hand,  seems to be a worthwhile exercise.

\acknowledgments 
We are grateful to  Augustinas Malinauskas, Priscilla Pani, Deborah Pinna and Giulia Zanderighi  for  useful discussions. 


\providecommand{\href}[2]{#2}\begingroup\raggedright\endgroup

\end{document}